\documentclass[a4paper,fleqn,usenatbib]{mnras}
\pdfoutput=1
\usepackage{mathptmx}
\usepackage[T1]{fontenc}
\usepackage{ae,aecompl}
\usepackage{graphicx}
\graphicspath{ {images/} }
\usepackage{amsmath}	
\usepackage{amssymb}	

\title[High-redshift dusty galaxies in HeLMS]{HerMES: A search for high-redshift dusty galaxies in the HerMES Large Mode Survey\,--\,Catalogue, number counts and early results}

\author[V. Asboth et al.]
{V. Asboth$^{1}$,
A. Conley$^{2}$,
J. Sayers$^{3}$,
M. B\'{e}thermin$^{4}$,
S. C. Chapman$^{5}$,
\newauthor
D. L. Clements$^{6}$,
A. Cooray$^{3,7}$,
H. Dannerbauer$^{8}$,
D. Farrah$^{9}$,
J. Glenn$^{2}$,
\newauthor
S. R. Golwala$^{3}$,
M. Halpern$^{1}$,
E. Ibar$^{10}$,
R. J. Ivison$^{11,4}$,
P. R. Maloney$^{2}$,
\newauthor
R. Marques-Chaves$^{13,14}$,
P. I. Martinez-Navajas$^{13,14}$,
S. J. Oliver$^{12}$,
\newauthor
I. P\'{e}rez-Fournon$^{13,14}$,
D. A. Riechers$^{15}$,
M. Rowan-Robinson$^{6}$,
Douglas Scott$^{1}$,
\newauthor
S. R. Siegel$^{3}$,
J. D. Vieira$^{16}$,
M. Viero$^{17}$,
L. Wang$^{18,19}$,
J. Wardlow$^{20,21}$,
J. Wheeler$^{2}$
\\
$^{1}$Department of Physics \& Astronomy, University of British Columbia, 6224 Agricultural Road, Vancouver, BC V6T-1Z1, Canada\\
$^{2}$Center for Astrophysics and Space Astronomy 389-UCB, University of Colorado, Boulder, CO 80309, USA\\
$^{3}$Division of Physics, Math, and Astronomy, California Institute of Technology, 1200 East California Blvd, Pasadena, CA 91125, USA\\
$^{4}$European Southern Observatory, Karl-Schwarzschild-Strasse 2, 85748 Garching, Germany\\
$^{5}$Department of Physics and Atmospheric Science, Dalhousie University, 6310 Coburg Road, Halifax, NS B3H 4R2, Canada\\
$^{6}$Astrophysics Group, Imperial College, Blackett Laboratory, Prince Consort Road, London, SW7 2AZ, UK\\
$^{7}$Center for Cosmology, Department of Physics and Astronomy, University of California, Irvine, CA 92697, USA\\
$^{8}$Universit\"{a}t Wien, Institut f\"{u}r Astrophysik, T\"{u}rkenschanzstrasse 17, 1180 Wien, Austria\\
$^{9}$Department of Physics, Virginia Tech, Blacksburg, VA 24061, USA\\
$^{10}$Instituto de F\'{i}sica y Astronom\'{i}a, Universidad de Valpara\'{i}so, Avda. Gran Breta\~na 1111, Valpara\'{i}so, Chile\\
$^{11}$Institute for Astronomy, Royal Observatory Edinburgh, Blackford Hill, Edinburgh, EH9 3HJ, UK\\
$^{12}$Astronomy Centre, Department of Physics \& Astronomy, University of Sussex, Brighton BN1 9QH, UK\\
$^{13}$Instituto de Astrof\'{i}sica de Canarias, E-38205 La Laguna, Tenerife, Spain\\
$^{14}$Universidad de La Laguna, Departamento de Astrof\'{i}sica, E-38206 La Laguna, Tenerife, Spain\\
$^{15}$Department of Astronomy, Cornell University, Space Sciences Building, Ithaca, NY 14853, USA\\
$^{16}$Department of Astronomy and Department of Physics, University of Illinois, 1002 West Green St., Urbana, IL 61801\\
$^{17}$Kavli Institute for Particle Astrophysics and Cosmology, Stanford University, 382 Via Pueblo Mall, Stanford, CA 94305, USA\\
$^{18}$SRON Netherlands Institute for Space Research, Landleven 12, 9747 AD, Groningen, The Netherlands\\
$^{19}$Institute for Computational Cosmology, Department of Physics, University of Durham, South Road, Durham, DH1 3LE, UK\\
$^{20}$Dark Cosmology Centre, Niels Bohr Institute, University of Copenhagen, Denmark\\
$^{21}$Centre for Extragalactic Astronomy, Department of Physics, Durham University, South Road, Durham, DH1 3LE, UK
}

\pubyear{2016}

\begin{document}

\label{firstpage}

\pagerange{\pageref{firstpage}--\pageref{lastpage}}

\maketitle

\begin{abstract}
Selecting sources with rising flux densities towards longer wavelengths from \textit{Herschel}/SPIRE maps is an efficient way to produce a catalogue rich in high-redshift (\textit{z} $>$ 4) dusty star-forming galaxies. The effectiveness of this approach has already been confirmed by spectroscopic follow-up observations, but the previously available catalogues made this way are limited by small survey areas. Here we apply a map-based search method to 274\,deg$^2$ of the HerMES Large Mode Survey (HeLMS) and create a catalogue of 477 objects with SPIRE flux densities $S_{500} > S_{350} >S_{250}$ and a $5 \, \sigma$ cut-off $S_{500} > 52$\,mJy. From this catalogue we determine that the total number of these ``red'' sources is at least an order of magnitude higher than predicted by galaxy evolution models. These results are in agreement with previous findings in smaller HerMES fields; however, due to our significantly larger sample size we are also able to investigate the shape of the red source counts for the first time. We have obtained spectroscopic redshift measurements for two of our sources using the Atacama Large Millimeter/submillimeter Array (ALMA). The redshifts \textit{z} = 5.1 and \textit{z} = 3.8 confirm that with our selection method we can indeed find high-redshift dusty star-forming galaxies.
\end{abstract}



\section{Introduction}

A key goal in studying the formation and evolution of galaxies is to understand how and when they formed the stars they contain. Although stars emit most of their radiation at UV and optical wavelengths, it has been shown that almost half of the extragalactic background radiation is made up of starlight absorbed and reprocessed by dust and re-emitted at far-infrared and submillimetre wavelengths \citep{puget1996,fixsen1998,devlin2009,leiton2015}. In the local Universe dusty starbursts with extreme infrared luminosities ($L_{\text{IR}}>10^{12} \text{L}_{\odot}$) are very rare, and contribute only a small fraction to the total star formation rate density. After the discovery of a higher redshift dusty galaxy population at submillimetre wavelengths with SCUBA \citep{smail1997,hughes1998,barger1998} it became clear that dusty star formation plays a significant role in the evolution of galaxies at earlier epochs. 

The contribution of the emission of dusty star-forming galaxies to the infrared luminosity density at different epochs has already been investigated up to $z \lesssim 4$ \citep[e.g.][]{gruppioni2013}. While several $z > 4$ luminous dusty star-forming galaxies have been detected \citep[e.g.][]{daddi2009,coppin2009,riechers2010,capak2011,walter2012,riechers2013,weiss2013,vieira2013}, their role in the stellar mass build-up at these high redshifts is still unknown. As of now, only a small set of observed fields have sufficient ancillary data coverage, limiting the samples available with known photometric redshift distributions. Additionally, dusty galaxies at $z>4$ are often undetectable at shorter wavelengths, thus the determination of the photometric redshifts becomes uncertain. To study the contribution of the high-redshift dusty galaxies to the star formation rate density we need a way to select large samples of $z>4$ objects based on their properties in the available far-infrared/submillimetre data sets alone.

\citet{dowell2014} constructed a sample of potentially $z > 4$ galaxies selected from  21\,deg$^2$ of data from the Herschel Multi-tiered Extragalactic Survey (HerMES, \citealp{oliver2012}) at wavelengths of 250, 350 and 500\,$\mu$m. The spectral energy distribution of local ultraluminous dusty star-forming galaxies typically peaks at rest-frame wavelengths $\lambda \sim 100$\,$\mu$m. At $z \gtrsim 4$ this peak shifts to wavelengths  $\lambda \gtrsim $ 500\,$\mu$m, thus all three SPIRE bands sample the short wavelength side of the SED peak of such high-$z$ objects, where the flux densities are increasing with $\lambda$. We note that not all high-redshift galaxies have red SPIRE colours, but this method provides a way to select predominantly $z>4$ objects based on their SPIRE flux densities alone. \citet{dowell2014} used a map-based search method to find sources with rising flux densities towards longer wavelengths ($S_{500} > S_{350} > S_{250}$). Follow-up observations proved that many of these sources are indeed at $z > 4$ (Riechers et al. in prep.), and this analysis resulted in the detection of the $z = 6.34$ source HFLS3 \citep{riechers2013,robson2014,cooray2014}, the highest redshift dusty starburst galaxy found to date, forming stars at a rate of several thousand solar masses per year. \citet{dowell2014} found an excess of these ``500-$\mu$m-riser'' or ``red'' objects compared to available galaxy evolution model predictions, and if the 10 or so red sources with spectroscopically confirmed high redshifts are representative of the whole population, then the number density of such galaxies poses a challenge to our current knowledge about galaxy evolution.

In this paper, as a continuation of the programme started by \citet{dowell2014}, we use a similar map-based search technique to create a large sample of 500-$\mu$m-riser galaxies by analysing a new field in the HerMES survey. The instrumental noise in this map is higher than the noise in any of the previously studied HerMES fields; however, the observed area is much larger than before; and therefore we find a statistically significant sample of brighter objects, including some strongly-lensed galaxies with flux densities above $S_{500}$ $ = 100$\,mJy as described by \citet{negrello2010}, \citet{paciga2009}, \citet{wardlow2013}, \citet{nayyeri2016} and others. 

In Section~\ref{sec:obs} we describe the observations of our new field. In Section~\ref{sec:cat} we present the steps used to create the catalogue. In Section~\ref{sec:dnds} we determine the raw number counts of our objects, examine biases in our source selection using Monte Carlo simulations, and compare our results to different models.  In Section~\ref{sec:tobs} we discuss the colour distribution, SED fits and apparent temperature distribution of our sample and in Section~\ref{sec:followup} we present results from follow-up observations of a sub-sample of our objects with ALMA and CSO/MUSIC. 
\section{Observations}
\label{sec:obs}

The HerMES Large Mode Survey (HeLMS) consists of a large area shallow observation of an equatorial field at wavelengths of 250, 350 and 500\,$\mu$m, obtained using the Spectral and Photometric Imaging Receiver (SPIRE, \citealp{griffin2010}) aboard the \textit{Herschel Space Observatory} \citep{pilbratt2010}. HeLMS is an extension of HerMES \citep{oliver2012}, a ``wedding cake'' type survey containing small and deep maps and larger shallower observations of different fields. HeLMS covers about 302\,deg$^2$ of the sky, making it the largest area observed in the HerMES survey. 

The HeLMS field spans $23^\text{h}14^\text{m} \textless \text{RA} \textless 1^\text{h} 16^\text{m}$ and $-9^\circ \textless \text{Dec} \textless +9^\circ$, an  equatorial region with low cirrus contamination. It was designed to have a large overlap with the Sloan Digital Sky Survey's Stripe 82 field \citep{abazajian2009}, one of the most highly observed areas of the sky, with extensive multi-wavelength ancillary data coverage. The equatorial area has the advantage that it can be observed from almost any ground-based telescope site in the world. 

The HeLMS field was observed with the telescope operating in fast-scan mode (60 arcsec s$^{-1}$ scanspeed). The observations were repeated in two nearly orthogonal scan directions, and these two data-sets are co-added during the map-making process. Having only two scans in each part of the map gives shallower coverage than the deepest SPIRE maps. However, the noise is still only a few times higher than the confusion level, and the large area of the survey allows us to find more of the rare objects contributing to the steep bright end of the number counts. 

\section{Catalogue creation}
\label{sec:cat}

We use a similar technique to find red sources in the HeLMS field as the map-based search method described in \citet{dowell2014}. In this method, instead of matching sources found independently at each wavelength, we smooth all maps to the same resolution, then we combine our observations at different wavelengths and use the information in the maps directly to find red sources. As a modification to the technique described in \citet{dowell2014}, we use a point source-matched filter instead of a Gaussian kernel to reduce the confusion noise in the smoothed maps.  

The map-based search method is better suited for finding red sources than catalogue based techniques. The currently available catalogue of HeLMS sources (Clarke et al., in prep.) uses the positions of the galaxies detected in the 250\,$\mu$m map as a prior to extract the flux densities at longer wavelengths, in order to reduce the effects of source blending. Thus this data set is not optimal for finding our typical red sources, since we expect these 500-$\mu$m-riser galaxies to have low signal-to-noise ratio in the 250\,$\mu$m maps, and hence many of them could go undetected in this catalogue. On the other hand, the \citet{dowell2014} sample already demonstrated that with a map-based method we can find sources that are not detected in various SPIRE catalogues that use shorter wavelength priors. 
 

\subsection{Maps}

\begin{figure*}
\centering
\includegraphics[width=0.8\textwidth]{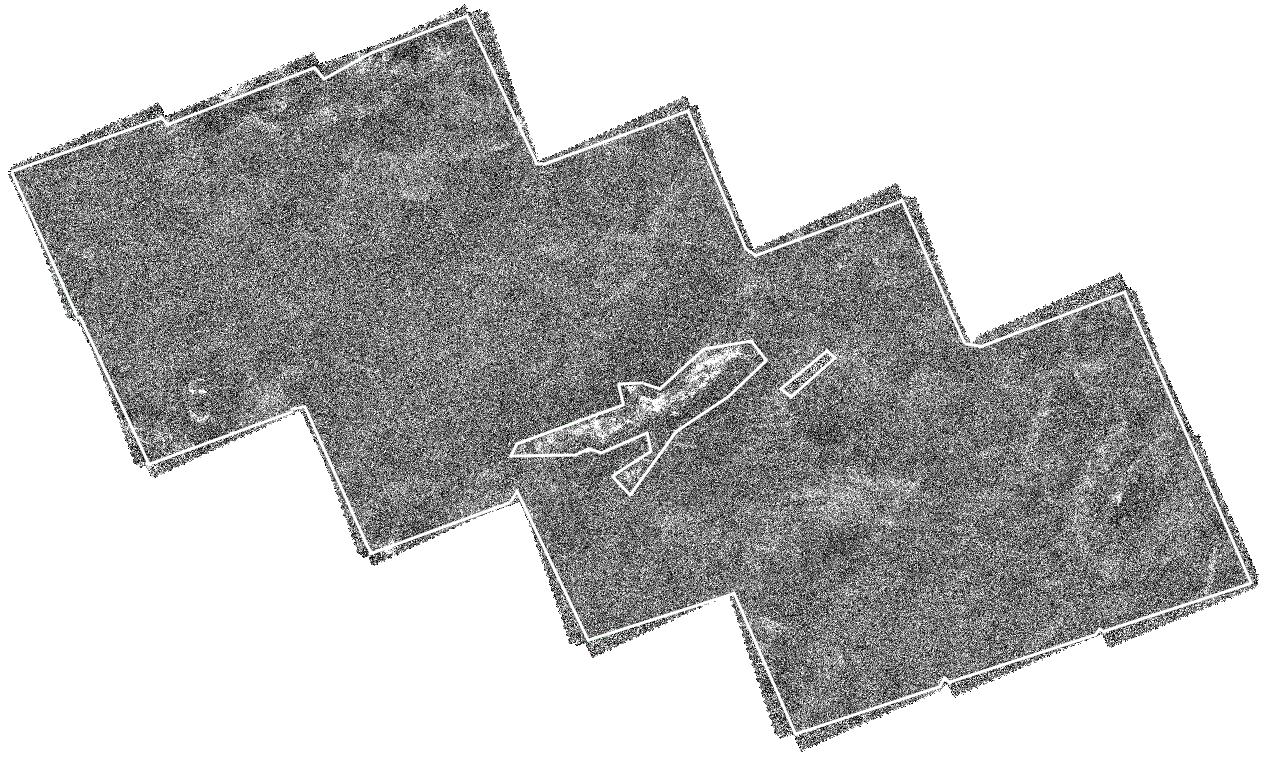} 
\caption{Greyscale image of the HeLMS 250\,$\mu$m map with solid lines showing the region we use in our analysis. The observed area spans about $30^{\circ}$ in RA and $ 18^{\circ}$ in Dec. We discard the edges of the maps, where the data lack overlapping scans, and we also discard a smaller region in the middle where part of the scan had to be removed and our coverage is sparse. We additionally mask out a ``seagull-shaped'' region of strong Galactic cirrus emission. The cirrus in this structure cannot be easily removed and biases our flux estimation of sources. The total area of the remaining data set after applying the mask is 273.9 deg$^2$. \label{fig:helmsarea}}.
\end{figure*}

We use the SMAP/SHIM iterative map-maker \citep{levenson2010} with the modifications described in \citet{viero2013} to create our maps. The nominal pixel sizes at 250, 350 and 500\,$\mu$m are $6''$, $8.333''$ and $12''$, respectively, to match one third of the full-width half-maximum (FWHM) of the beam in each band ($18''$, $25''$, $36''$). Since we want to combine our observations, we create all three of our maps with matching pixel sizes of $6''$ instead. 

We discard the edges of the map, where the telescope turned around between scans and the data are not cross-linked. This area is too noisy and the coverage is too sparse to reliably estimate the fluxes of our objects. Similarly, we discard a small region in the middle of the map, where part of one of the overlapping scans had to be removed due to stray light in the telescope. The large-scale cirrus background is subtracted during the source-finding method, but there is a ``seagull-shaped'' area in the middle of the maps, where the cirrus is too strong to be easily removed and the flux estimations are biased high, so we mask this region manually (see Fig.~\ref{fig:helmsarea}). The total remaining area that we use in our analysis is 273.9\,deg$^2$. 
 
\subsection{Matched filter}
\label{sec:mfilt}

\begin{figure}
\centering
\includegraphics{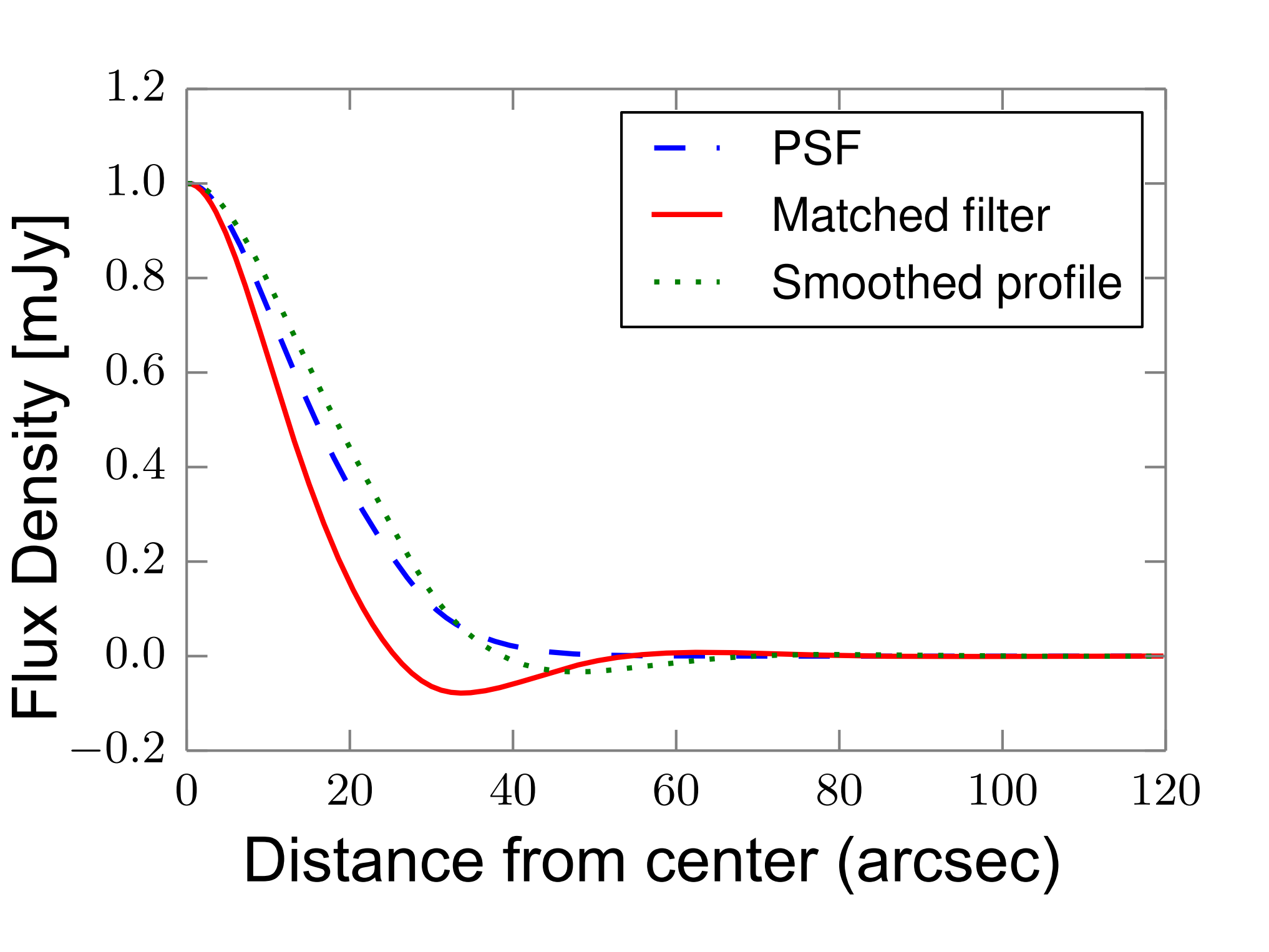} 
\caption{Shape of the matched filter at 500\,$\mu$m, compared to the Gaussian point spread function (PSF) and the final source profile in the smoothed maps. The filter used for smoothing our maps has a smaller full-width half-maximum than the beam, so the resolution of our maps after smoothing is close to the unsmoothed resolution.\label{fig:mfilt}}
\end{figure}

\begin{figure}
\centering
\includegraphics[width=.40\textwidth]{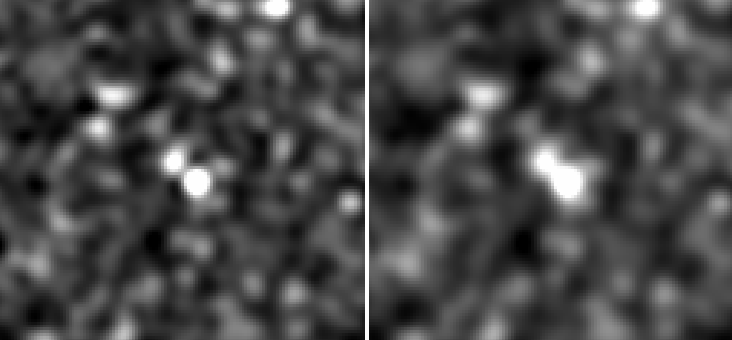} 
\caption{Comparison of the matched-resolution maps when using different smoothing kernels. On the left, we show a $13' \times 13'$ cut-out image from our 250\,$\mu$m map smoothed with the matched filter described in Section~\ref{sec:mfilt}, while on the right we used a Gaussian filter. In addition to reducing the confusion noise from unresolved faint sources in our telescope beam, the matched filter also reduces blending effects between neighbouring bright sources. \label{fig:filt_diff}}
\end{figure}

To create maps with matching resolution we use an optimal filter that maximizes the signal-to-noise ratio in a map with non-negligible confusion noise. This filter is described in detail in  \citet{chapin2011}. The signal-to-noise ratio in Fourier space after we cross-correlate our signal $S$ with our filter $F$ is

\begin{equation}
\text{SNR} = \frac{\sum_k \hat{F}_k^\text{T} \hat{S}_k}{\left( \sum_k {\lvert \hat{F}_k^\text{T} \hat{N}_k \rvert}^2 \right)^{1/2} }\text{.}\end{equation}
Here $N$ is the total noise, including instrumental noise and confusion, the hats denote the discrete Fourier transforms of our variables, the ``T'' superscript refers to a transpose of our filter, and the index $k$ corresponds to components in the spatial frequency domain. We can derive the optimal filter by finding $F$ for which 
\begin{equation}\frac{\partial \text{(SNR)} }{\partial \hat{F}_j^\text{T}} = 0 \text{.}\end{equation}
The resulting filter is
\begin{equation}\hat{F}_j^\text{T} = \frac{\hat{S}_j}{{\lvert \hat{N}_j \rvert}^2} \left( \frac{\sum_k {\lvert \hat{F}_k^\text{T} \hat{N}_k \rvert}^2}{\sum_k \hat{F}_k^\text{T} \hat{S}_k} \right)\text{,}\end{equation}
where $N_j$ represents the total noise at each frequency component $j$. While the instrumental noise is white and its value is constant at all frequencies, the power spectrum of the confusion noise will have a similar shape to the point spread function, since confusion arises from point sources in the same beam. The shape of our matched filter can be seen in Fig.~\ref{fig:mfilt}, compared to the 500\,$\mu$m beam shape and the final source profile after applying the filter to our maps. Due to the smaller width of our filter, our final resolution will be closer to the original resolution of the 500\,$\mu$m map than in \citet{dowell2014}, and this can help reduce source blending effects for nearby objects (see an example in Fig.~\ref{fig:filt_diff}).

After finding the optimal matched filter for the 500\,$\mu$m map, we need to construct the smoothing kernels $K_{250,350}$ that create the same effective source shape at 250 and 350\,$\mu$m that we measure in the smoothed 500\,$\mu$m map. First we convolve the 500\,$\mu$m  beam (a Gaussian with $35.3''$ FWHM) with the matched filter to find the final source shape in our smoothed maps. If $P_{250,350}$ denotes the nominal beam shapes at 250 and 350\,$\mu$m and $P_{\text{mf}500}$ is the matched-filtered source-shape at 500\,$\mu$m, then we can find $K_{250,350}$ from the convolution
\begin{equation} P_{250,350} \otimes K_{250,350} = P_{\text{mf}500} \text{.}\end{equation}
Thus the smoothing kernels are determined by taking the inverse Fourier transformation of the Fourier-space ratio of the final and initial beam shapes.

Before filtering the maps, we subtract a local background constructed by smoothing our maps with a 2D median boxcar filter on a $3'$ scale, to remove any large-scale cirrus fluctuations that might otherwise affect our flux estimation. The SPIRE maps contain an error extension, which is an output of the mapmaker. Each pixel in this error map contains the standard deviation of the flux density values from the time-ordered detector data that are projected onto that given pixel. We use the inverse-variance values calculated from these error values as weights for each pixel when we filter our data with the matching kernels. Since the mapmaking pipeline does not correct for the effects of pixelization, we create our filters on an oversampled grid, and then rebin them to our final pixel size. We also apply our filters to our error maps to find the typical instrumental noise values in our pixels after smoothing. We test this filtering method by injecting fake sources with known flux density values into our raw maps, and find no significant bias in the recovered flux distribution after subtracting the background and applying our filter.

\subsection{Difference map}

Because the sources responsible for the confusion noise in the maps emit at all three SPIRE wavelengths, one can produce a difference map that has a substantially reduced confusion limit. It will be much more effective to search for bright 500\,$\mu$m sources in such a difference map than in the raw 500\,$\mu$m flux maps. \citet{dowell2014} found that the difference 
\begin{equation}D = \sqrt{1 -k^2} M_{500} - k M_{250}\end{equation}
reduces confusion, while red sources remain bright in the $D$ map. While the confusion noise is strongly correlated between the SPIRE bands, the instrumental noise is uncorrelated between maps, and the arbitrary normalization was chosen so that the instrumental noise properties of the difference map would not change if the noise values in the original maps are comparable. The coefficients were optimized by investigating the efficiency of recovering artificial red sources injected into simulated sky maps. \citet{dowell2014} demonstrated that the value $k = 0.392$ works well empirically to maximize $D / \sigma_{\text{conf}}$ in the maps. They also experimented with creating linear combinations using all three maps, but they found that including a 350\,$\mu$m term does not improve the efficiency of the source selection. We find that this same choice of coefficients also works well for our HeLMS maps. Our final difference map is constructed as 
\begin{equation}
\label{eq:diffmap}
 D = 0.92M_{500}-0.392M_{250}. 
\end{equation}
We also combine the error maps and determine the resulting instrumental noise ($\sigma_{\text{instr}}$) levels. After measuring the total variance ($\sigma^2_{\text{total}}$) of our map we calculate the confusion noise ($\sigma_{\text{conf}}$) in our final map as 

\begin{equation}\sigma_{\text{conf}}=\sqrt{\sigma^2_{\text{total}} - \sigma^2_{\text{instr}}} \text{.}\end{equation}
The noise levels in our smoothed maps and the difference map are listed in Table ~\ref{tbl:noise}. It is clearly seen that the confusion noise in the difference map is reduced compared to the confusion levels in the single band maps.

\begin{table}
\centering
\caption{Noise levels in the maps \label{tbl:noise}}
\begin{tabular}{crrr}
\hline
  & \multicolumn{1}{c}{$\sigma_{\text{tot}}$} & \multicolumn{1}{c}{$\sigma_{\text{inst}}$}  & \multicolumn{1}{c}{$\sigma_{\text{conf}}$} \\ 
  & [mJy] & [mJy] & [mJy] \\ 
\hline
250\,$\mu$m & 15.61 & 7.56 & 13.66 \\ 
350\,$\mu$m & 12.88 & 6.33 & 11.21 \\ 
500\,$\mu$m & 10.45 & 7.77 & 6.98 \\ 
\textit{D} & 8.54 & 7.75 & 3.50 \\ 
\hline
\end{tabular} 

\begin{flushleft}
\footnotesize \textbf{Note.} -- The 1 $\sigma$ total, instrumental and confusion noise levels in our smoothed 250, 350 and 500\,$\mu$m maps and in the difference map.
\end{flushleft}

\end{table}

\subsection{Source extraction}

\begin{table*}
\caption{List of the ten brightest high-redshift dusty galaxy candidates in HeLMS.  \label{tbl:redlist}}
\begin{tabular}{lrrrrrrrr}
\hline
\multicolumn{1}{c}{Source Name} & \multicolumn{1}{c}{RA} & \multicolumn{1}{c}{Dec} & \multicolumn{1}{c}{$S_{\text{250}}$} & \multicolumn{1}{c}{$r_{\text{250}}$} &\multicolumn{1}{c}{$S_{\text{350}}$} & \multicolumn{1}{c}{$r_{\text{350}}$} &\multicolumn{1}{c}{$ S_{\text{500}}$} & \multicolumn{1}{c}{$ r_{\text{500}}$} \\
 & \multicolumn{1}{c}{[deg]} & \multicolumn{1}{c}{[deg]} & \multicolumn{1}{c}{[mJy]} & & \multicolumn{1}{c}{[mJy]} & & \multicolumn{1}{c}{[mJy]} & \\
\hline
HELMS{\_}RED{\_}1 & 11.0414 & 1.3063 & $ 108.1 \pm   6.9 $ &   0.89 & $ 166.5 \pm   6.0 $ &   0.89 & $ 191.8  \pm   8.2 $ &   0.90 \\ 
HELMS{\_}RED{\_}2 & 13.2455 & 6.2219 & $  68.2 \pm   6.0 $ &   0.77 & $ 111.6 \pm   5.9 $ &   0.93 & $ 131.7  \pm   6.9 $ &   0.94 \\ 
HELMS{\_}RED{\_}3 & 9.8731 & 0.4067 & $ 140.8 \pm   6.5 $ &   0.93 & $ 152.6 \pm   6.3 $ &   0.88 & $ 162.1  \pm   7.3 $ &   0.93 \\ 
HELMS{\_}RED{\_}4 & 5.5867 & $-1.9224$ & $  62.2 \pm   6.1 $ &   0.83 & $ 104.1 \pm   5.8 $ &   0.87 & $ 116.4  \pm   6.6 $ &   0.89 \\ 
HELMS{\_}RED{\_}5 & 12.6982 & 6.9556 & $  20.8 \pm   6.0 $ &   0.46 & $  68.2 \pm   6.4 $ &   0.71 & $ 112.0 \pm   6.8 $ &   0.80 \\ 
HELMS{\_}RED{\_}6 & 15.2248 & 3.0563 & $  50.1 \pm   6.8 $ &   0.72 & $  83.3 \pm   6.1 $ &   0.88 & $  96.1  \pm   7.8 $ &   0.83 \\ 
HELMS{\_}RED{\_}7 & 9.5584 & $-0.3806$ & $  73.4 \pm   5.6 $ &   0.79 & $ 119.0 \pm   6.0 $ &   0.89 & $ 122.9  \pm   6.7 $ &   0.87 \\ 
HELMS{\_}RED{\_}8 & 354.5083 & $-1.3186$ & $  33.6 \pm   6.5 $ &   0.56 & $  53.8 \pm   6.1 $ &   0.69 & $  90.9  \pm   7.6 $ &   0.82 \\ 
HELMS{\_}RED{\_}9 & 6.8253 & 2.6629 & $  65.2 \pm   5.9 $ &   0.79 & $  76.4 \pm   5.7 $ &   0.88 & $  99.3  \pm   6.9 $ &   0.91 \\ 
HELMS{\_}RED{\_}10 & 0.7683 & 2.6865 & $  33.6 \pm   5.7 $ &   0.67 & $  53.9 \pm   6.5 $ &   0.77 & $  86.5  \pm   7.0 $ &   0.86 \\ 
\hline
\end{tabular} 
\begin{flushleft}
\footnotesize \textbf{Note.} -- The full catalogue of 477 red sources is available in the contributed data section of HeDaM/HerMES (http://hedam.lam.fr/HerMES/).
\end{flushleft}
\end{table*}

To find red sources in our maps we first search for the brightest peaks in our minimum variance difference map and then we select the 500-$\mu$m-riser objects from the resulting list. We apply a local-maxima search algorithm to our \textit{D} map, which finds the positions of the pixels that have  greater values than their eight adjacent pixels. We create a list of these peaks with a cut-off at $D = 34$\,mJy which corresponds to 4$\sigma_{\text{tot},D}$ in the difference map. This cut is determined by simulating the number of false-detections in the SPIRE maps. We create simulated maps from mock catalogues drawn from the \citet{bethermin2012} model, but we remove any intrinsically red sources from these catalogues. Then we run our red-source detection pipeline on these simulated maps, and we measure the number of objects that we detect as a red source in the final maps. If our real catalogue contains $N_\text{cat}$ sources and in a simulated map we detect $N_\text{false}$ intrinsically non-red objects with red measured colours, then we can determine the purity of our catalogue as $p= (N_\text{cat} - N_\text{false})/N_\text{cat}$. Our simulations show that the purity of our HeLMS red source catalogue decreases from 94$\%$ at $D > 4 \sigma_{D}$ to 77$\%$  if we apply a $D > 3  \sigma_{D}$ cut. 

To select red sources from the list of peaks found in the $D$-map, we simply require that $S_{500} > S_{350} > S_{250}$. However, to evaluate this we need to extract the actual flux densities from our single wavelength maps at these \textit{D}-peak positions. It is not trivial to determine if it is optimal to use our smoothed maps to measure these values or to go back to the nominal resolution maps and find the sources there. From simulations we know that we have a typical positional uncertainty of $6''$, so extracting the fluxes at the precise \textit{D} position biases our flux estimation at 250 and 350\,$\mu$m. To address this we could re-fit our peaks in each of the smoothed maps to find the actual peak position in each band and extract the fluxes there. However, a typical red source has an $S_{500}$/$S_{350}$ flux density ratio that is close to 1, and hence adjacent sources often boost our 350\,$\mu$m flux density above $S_{500}$, even if in the nominal maps we clearly detect our source as a red source. This is an important issue at the bright end, where the source counts decrease rapidly, and even in our very large area field we expect only to find a handful of such objects. After careful consideration, we decided that for this last step it is better to measure the fluxes from the less confused nominal resolution maps, but instead of performing photometry at the measured $D$-map position, we find the best-fit source after taking into account our positional uncertainty. 

To achieve this we move around our \textit{D}-map peak position in sub-pixel steps, allowing the search radius to change, corresponding to our typical uncertainty, and we calculate the Pearson correlation coefficient $r$ between our data $d$ and the beam shape $P$ at each position by
\begin{equation} \label{eq:corr}
r =  \frac{\displaystyle \sum_{i=1}^{N_{\text{pixels}}}(d_i - \bar{d})(P_i - \bar{P})}{\displaystyle \left[\sum_{i=1}^{N_{\text{pixels}}}(d_i - \bar{d})^2\right]^{1/2}\left[\sum_{i=1}^{N_{\text{pixels}}}(P_i - \bar{P})^2\right]^{1/2}} \text{.}\end{equation}
We pick the position where the correlation is the largest, and we extract the flux density at this position using inverse variance weighting:
\begin{equation}S =\frac{\displaystyle \sum_{i=1}^{N_{\text{pixels}}} d_i P_i / \sigma_i^2}{\displaystyle \sum_{i=1}^{N_{\text{pixels}}} P_i^2 / \sigma_i^2} \text{.} \end{equation}

We test the validity of this method by injecting artificial sources with known flux densities in the raw maps, and we run our source extraction pipeline on these maps. We find that this method reduces the bias due to positional uncertainties. However we note that this method will bias our flux estimation at 250 and 350\,$\mu$m wavelengths if our sources clearly break up into multiple components in these higher resolution maps. The only requirement we impose on our catalogue is that each object is detected as a point source at the 500\,$\mu$m resolution and appears red in the maps, but we do not require a clear high signal-to-noise detection in 250 and 350\,$\mu$m in order not to bias against the reddest objects. This means that source blending will be an issue in our catalogue, and we will discuss the effects of blending in the following section. We include the correlation values in our catalogue to show how reliable our flux density estimation is in each band.

Our catalogues could be contaminated by cosmic ray hits or other spikes in the detector timestreams that are not properly removed during data processing. A spike left in the 500\,$\mu$m array data would mimic a 250\,$\mu$m dropout source. During the iterative mapmaking process for ordinary SPIRE data, these residual isolated spikes are recognized as outliers among the samples associated with a given pixel, and are removed from the data \citep{viero2013}. However, the HeLMS maps are sparsely sampled and there may be too few
samples near a given pixel for this recognition procedure to be reliable. The result is a ``hot'' pixel or a stripe of a few very bright pixels in the map from one array, while the neighbouring pixels show values consistent with the instrumental noise and no spike is present in the other arrays. After smoothing the map with our matched filters, these corrupted pixels appear like bright sources in the 500\,$\mu$m map.

A common method to detect these objects is to create two ``jackknife'' maps, each from one half of the data. The false sources only show up in one of the maps. However, the very shallow depth of our observation causes these half-maps to be even more sparse, and so they contain new artefacts due to the lack of cross-linking. In the case of the HeLMS observations this method does not reliably remove cosmic rays from the maps. Instead we turn to a different approach. For each source we compare the raw and smoothed maps in a $5 \times 5$ pixel region around each source, and discard all candidate sources if any pixel shows a large difference, $(S_{\text{raw}}-S_{\text{smooth}}) > 5 \sigma_{\text{raw}}$. Most cosmic rays produce outlier pixels almost 10 $\sigma$ away from the smoothed values, and this method works well to discard these false sources. In our final catalogue we adopt a 5$\sigma$ cut of $S_{500}$ $> 52$\,mJy to protect from fainter cosmic rays that this technique may not have recognized.

Flat spectrum radio quasars at $z < 1$ can have similar colours to those of our high-redshift galaxy candidates, but these objects can be easily identified from available radio surveys. We compared our catalogue to the 21-cm radio catalogues from the NRAO VLA Sky Survey (NVSS, \citealp{condon1998}) and the FIRST survey \citep{becker1995} and we discarded 17 of our sources from our catalogue that have radio flux densities brighter than 1\,mJy. We do not use these objects in  any further analysis. After discarding these radio quasars, our final catalogue contains 477 red sources with $S_{500} > S_{350} >S_{250}$, $S_{500}$ $>52$\,mJy and $D>34$\,mJy. The positions, measured SPIRE flux densities and the $r$ Pearson correlation coefficients (see equation~\ref{eq:corr}) of our ten brightest sources are listed in Table~\ref{tbl:redlist}. The full catalogue of 477 red sources is available online in the contributed data section of HeDaM/HerMES (http://hedam.lam.fr/HerMES/). 

\section{Number counts}
\label{sec:dnds}

\begin{figure*}
\centering
\includegraphics[width=0.8\textwidth]{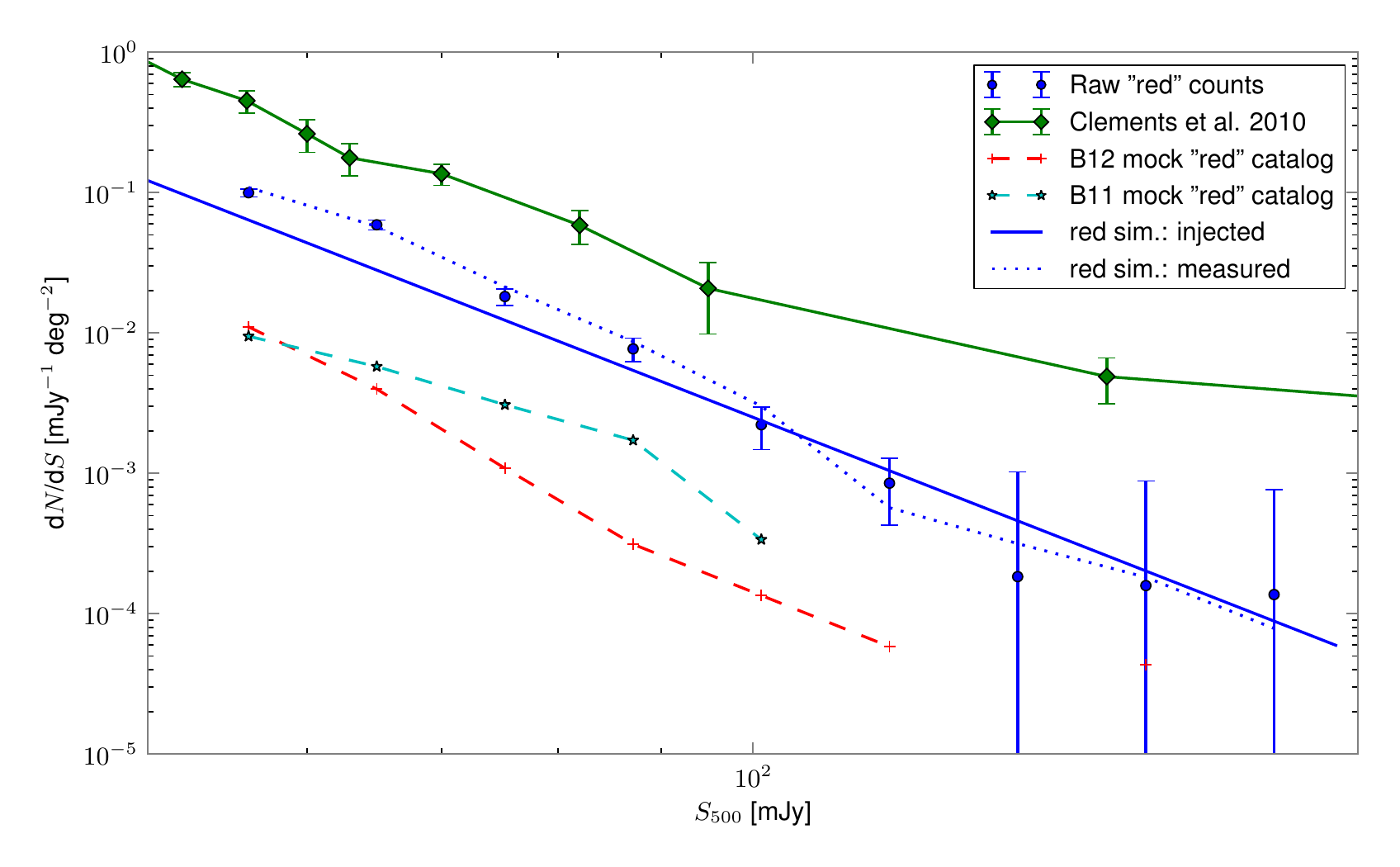}
\caption{Raw 500\,$\mu$m differential number counts of our sample of ``red'' sources. Filled circles represent the raw red source counts with 1$\sigma$ Poisson error bars, except for the highest three flux density bins, where we plot the 95$\%$ upper confidence limits. The green diamonds show the total \textit{Herschel} 500\,$\mu$m number counts measured by  \protect\citet{clements2010}. The dotted blue line represents measured raw counts from a 1500 deg$^2$ simulation, where artificial red sources were drawn from an intrinsic red source distribution shown by the solid blue line and these sources were injected into simulated sky-maps containing instrumental and confusion noise. These simulations that account for blending, Eddington bias, false detections and completeness are described in Section~\ref{sec:sims}. The light-blue stars connected with a dashed line are binned data from creating a simulated catalogue based on the \protect\citet{bethermin2011} model, and selecting objects with the same colour criteria as we do for our catalogue. The red crosses connected with a dashed line show the number counts from a similarly selected mock catalogue drawn from the \protect\citet{bethermin2012} model. The model comparisons are discussed in Section~\ref{sec:models}. \label{fig:dnds}}
\end{figure*}

We measure the raw 500\,$\mu$m  differential number counts of the red sources in our final catalogue. The uncorrected numbers are shown in Table \ref{tbl:dnds} and plotted in Figure~\ref{fig:dnds}. In practice raw number counts need to be corrected for completeness and flux-boosting effects and the expected number of false detections needs to be subtracted from the binned data in order to infer the underlying true source distribution. When measuring number counts at a single wavelength these corrections usually only depend on the flux density and signal-to-noise ratio, and they are relatively easy to simulate when we are investigating sources with flux densities far above the confusion limit. Our catalogue, however, has a more complicated selection function, and the correction factors will also depend on the colours of our sources, with these colours having a very large scatter due to the low signal-to-noise ratio at our shorter wavelengths (see error bars in Fig.~\ref{fig:colourcolour}). Additionally, these corrections require us to assume an intrinsic shape for our number counts based on model predictions and previous observations; however, due to the small sample sizes the slope of the red source counts has not been measured before. Here we do not explicitly correct our estimated counts for these biases, but instead, in the next subsection, we describe a simulation where we attempt to predict the most likely shape of our intrinsic source counts, while taking the biasing effects into account.\\

\begin{table}
\caption{Raw 500\,$\mu$m number counts. \label{tbl:dnds}}
\begin{tabular}{ccccc}
\hline
\multicolumn{1}{c}{$S_{\text{min}}$} & \multicolumn{1}{c}{$S_{\text{max}}$} & \multicolumn{1}{c}{$S_{\text{mean}}$} & \multicolumn{1}{c}{$N_{\text{bin}}$}  &  \multicolumn{1}{c}{d$N$/d$S$} \\
  \multicolumn{1}{c}{[mJy]}   & \multicolumn{1}{c}{[mJy]}     & \multicolumn{1}{c}{[mJy]}    &            & \multicolumn{1}{c}{ [$\times 10^{-4}$ mJy$^{-1}$deg$^{-2}$]}  \\ 
\hline
 52.0     & 60.2      & 56.1       & 225        & $998.6 \pm 66.6$\\ 
 60.2     & 69.8      & 65.0       & 154        & $590.1 \pm 47.5 $\\ 
 69.8     & 80.8      & 75.3       & 55         & $181.9 \pm 24.5 $ \\ 
 80.8     & 93.6      & 87.2       & 27         & $\phantom{1}77.1 \pm 14.8 $ \\ 
 93.6     & 108.4     & 101.0      & 9          & $\phantom{1}22.2 \pm \text{ }7.4 $ \\ 
108.4     & 125.5     & 117.0      & 4          & $\phantom{1}\phantom{1}8.5 \pm \text{ }4.3 $ \\ 
125.5     & 145.4     & 135.4      & 1          & $\phantom{1}1.84\substack{\text{ }+\text{ }8.41 \\ \text{ }-\text{ }1.79}$\\ 
145.4     & 168.4     & 156.9      & 1          & $\phantom{1}1.58\substack{\text{ }+\text{ }7.26 \\ \text{ }-\text{ }1.55}$\\ 
168.4     & 195.0     & 181.7      & 1          & $\phantom{1}1.37\substack{\text{ }+\text{ }6.27 \\ \text{ }-\text{ }1.33}$\\ 
\hline
\end{tabular} 
\end{table}

\subsection{Simulations}
\label{sec:sims}

The raw source counts can provide a biased estimator of the intrinsic source distribution through a number of effects.  If the intrinsic source counts are a steep function of source brightness, or the flux uncertainties are large, a larger number of faint sources may appear to satisfy our catalogue selection criteria than the number of acceptable sources that appear not to. This so-called ``Eddington bias'' may be present at the cuts where we require  $S_{500}>52$\,mJy,  $D>34 $\,mJy, $S_{500}>  S_{350}$ and $S_{350} > S_{250}$, and this will affect the observed slope of the number counts.  

\begin{figure}
\centering
\includegraphics[width=.48\textwidth]{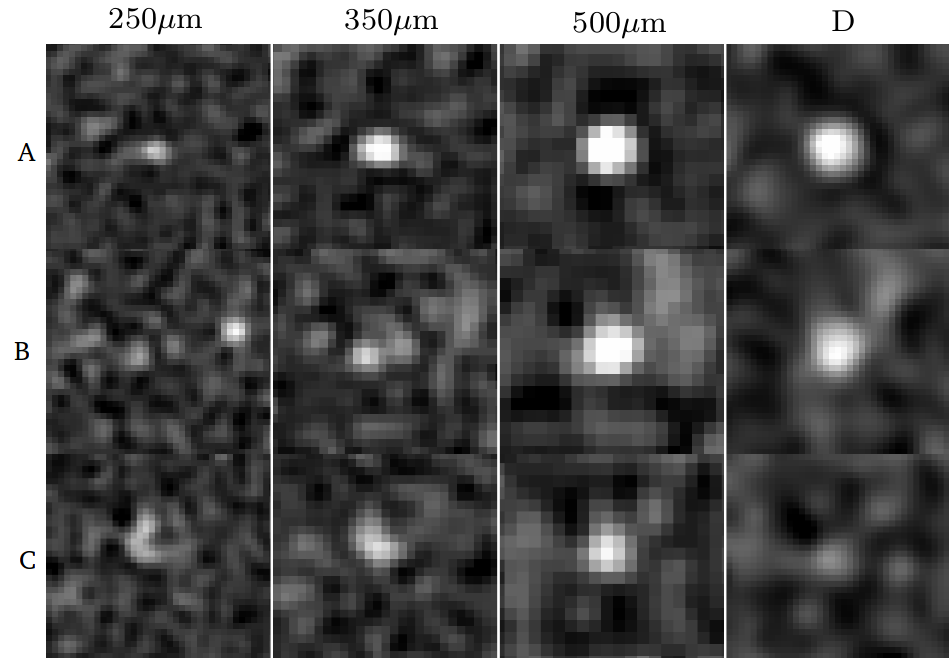} 
\caption{Example images of sources detected in our catalogue that appear to be a single object in the 500\,$\mu$m map and in \textit{D}. From left to right we show $3.5' \times 3.5'$ cut-out images at the nominal resolution of the 250\,$\mu$m, 350\,$\mu$m and 500\,$\mu$m maps and the difference map, respectively. Example A shows a source that is isolated and has a clear counterpart in each band. Example B shows an object that clearly breaks up into two sources in the 250\,$\mu$m and 350\,$\mu$m maps, but is blended at 500\,$\mu$m. Example C shows a complex blend, with no clearly identifiable counterparts that could be used for deblending. \label{fig:blends}}
\end{figure}

Additionally, there may be a bias in our counts that arises from the variation of our angular resolution with wavelength. There will be closely adjacent sources that appear blended into one object at 500\,$\mu$m and resolved into several at 350 or 250\,$\mu$m. If one of these sources is very red, but not bright enough for catalogue inclusion, and the other(s) neither very bright nor very red, the sum may well appear both bright enough and red enough for inclusion. This would result in a fairly red but slightly faint object in the catalogue. Examples of this effect are shown in Figure~\ref{fig:blends}.

Ordinarily these effects can be estimated and corrected using a model or prior knowledge of the shape of the intrinsic source counts. However, as discussed in \citet{dowell2014}, the existing models appear to under-predict the abundance of these sources by an order of magnitude or more, so this procedure will not be reliable.  Instead, we take a self-consistent approach. We use existing source models to construct an artificial sky without red sources, add a power law distribution of red sources and vary the terms in the power law until searches of the artificial maps return counts that match the raw counts in Fig.~\ref{fig:dnds}.

\begin{figure}
\centering
\includegraphics{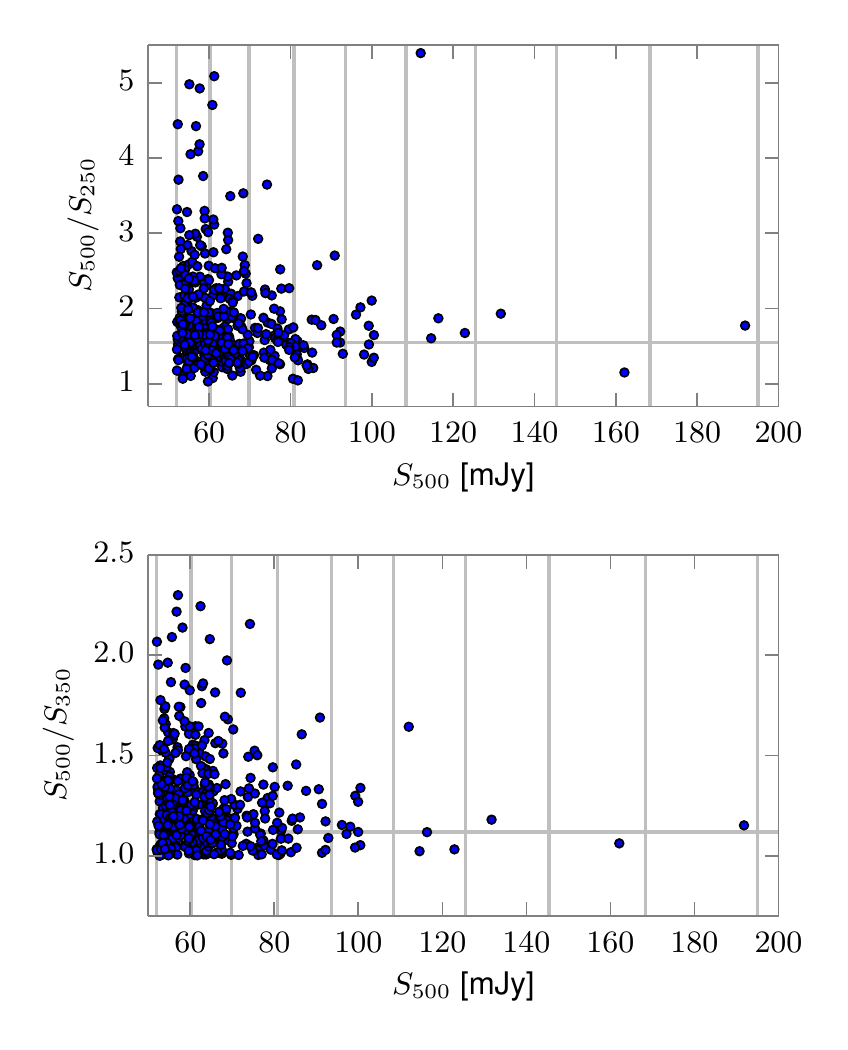} 
\caption[mf list of fig]{Measured SPIRE colours of red sources in our catalogue as a function of the 500\,$\mu$m flux density. The vertical lines show the edges of the flux density bins we used to measure the differential number counts. The horizontal lines at $S_{500}/S_{250} = 1.55$ and $S_{500}/S_{350} = 1.12$ show the median colours of our sample. The sources with anomalously red colours are mostly blends or objects without clear 250 and 350\,$\mu$m detections. \label{fig:coloursperbin}}
\end{figure}

We use the \citet{bethermin2012} model to create simulated maps at all three wavelengths. We discard any red sources in these maps, then we add the measured instrumental noise and we inject artificial red sources into this data set. To do this we first draw 500\,$\mu$m flux densities from a power-law distribution of the shape $dN/dS = N_0 \times S^{-\alpha}$, then we fix the colour ratios of our injected objects to the median of the colour ratios measured in our catalogue, $S_{500}/S_{250}$ = 1.55 and $S_{500}/S_{350}$ = 1.12. In Fig.~\ref{fig:coloursperbin} we plot the measured colour distribution in our 500\,$\mu$m bins. The largest scatter is in the lowest two bins, where many of the objects with high colour ratios are either blends or have very low signal-to-noise ratio counterparts at 250 and 350\,$\mu$m. After injecting the simulated population of red sources into the maps, we run our detection pipeline in the same way as we do for our real maps and we compare the measured counts from this simulated data with our raw numbers. In an iterative process we change the input parameters $N_0$ and $\alpha$ until the output counts are within the uncertainties compared to the measured raw number counts in our real data. In Fig.~\ref{fig:dnds} we show the result of a simulation with an area of 1500 deg$^2$. The blue solid line represents a power law $dN/dS = N_0 \times S^{\alpha}$ with $\alpha = -5.60$ and $\log N_0 = 8.61$ and the dotted blue line shows the measured number counts for simulated objects with 500\,$\mu$m flux densities drawn from this distribution and with fixed colour ratios. The simulations are in good agreement with our observed counts, suggesting that assuming a power-law number counts model without a break is adequate for our purposes. The parameters in our simulation have typical uncertainties of $\pm 0.01$; changing  $\alpha$ and $\log N_0$ by more than this value results in simulated counts outside of the $\pm 1\sigma $ error bars of our measured red source counts. We note that the actual colours of our objects are not all the same, but due to the large uncertainties in the colour measurements it is not straightforward to determine the underlying colour distribution.

\subsection{Comparison to models}
\label{sec:models}

We compare our observed number counts to mock ``red'' catalogues created from the \citet[B11]{bethermin2011} and \citet[B12]{bethermin2012} models. We generate 1000\,deg$^2$ simulations from both of these catalogues, and then select sources the same way as we do for our observed sample ($S_{500}> 52$\,mJy, $D>34$\,mJy, and ``red'' $S_{500} > S_{350} >S_{250}$ colours). These two models describe the total \textit{Herschel} number counts well, but, as \citet{dowell2014} already showed, they both under-predict the number of red sources in the HerMES fields. The resulting counts from these simulations are plotted in Fig.~\ref{fig:dnds}. Comparing these simulated number counts to our inferred intrinsic source distribution, we can see that both of these models indeed under-predict the number of red sources by at least an order of magnitude. We note that both of these models are empirical and are based on extrapolation of the properties of lower redshift starburst galaxies, such as the luminosity function (B11) or the stellar mass function (B12), instead of the actual physics of these galaxies. As of now, none of the more physically motivated models have been fine-tuned to properly describe the observed \textit{Herschel} number counts, hence those models are even less optimal to predict the red counts.

In Fig.~\ref{fig:dnds} we also show the total SPIRE 500\,$\mu$m counts measured by  \citet{clements2010}, which is in good agreement with other SPIRE total number counts measurements \citep{oliver2010,glenn2010, bethermin2012b}. \citet{negrello2010} predicted that the number counts of unlensed galaxies at 500\,$\mu$m are rapidly decreasing and reach zero at $>$ 100\,mJy. Many bright objects -- that are not a local galaxy or a blazar -- with flux densities above $S_{500} \simeq 100$\,mJy are expected to be strongly lensed, and the number counts are expected to have a bright tail due to this lensed population. We have found several red objects with bright 500\,$\mu$m flux densities, but their low numbers in our flux density bins do not allow us to see an actual departure from the steep power-law shape of red sources at lower flux densities. We note that six red galaxies in our sample (HELMS{\_}RED{\_}1, 2, 3, 4, 7, 11) are also present in the list of HeLMS lensed galaxy candidates with $S_{500} > 100$\,mJy discussed in \citet{nayyeri2016}, and five of our sources (HELMS{\_}RED{\_}1, 3, 4, 7 and 23) are part of the nine candidate gravitationally-lensed dusty star-forming galaxies detected using the Atacama Cosmology Telescope (ACT) at a wavelength of 1.4 mm \citep{su2015}. 

Our inferred number counts do not show any significant differences compared with the \citet{dowell2014} findings, although we now have much better statistics . The \citet{dowell2014} sample consisted of red sources with $S_{500} > 30$\,mJy and $D >24$\,mJy, and they found the total corrected cumulative counts to be $3.3 \pm 0.8$ sources per deg$^2$. If we assume that the shape of our distribution can be described by the power law that we found in Section~\ref{sec:sims}, and that it does not have a break between 30\,mJy and 52\,mJy, then integrating this power law above 30\,mJy gives us a total number of more than 10 red sources per deg$^2$. However, this comparison is not straightforward, since many of those objects at lower 500\,$\mu$m flux densities would be discarded by the \textit{D} cut used in the \citet{dowell2014} analysis. The $D >24$\,mJy cut imposes a maximum limit of $S_{250} < 9.2$\,mJy for an object with $S_{500} = 30$\,mJy (see Eq.~\ref{eq:diffmap} ).  At $S_{500}= 40$\,mJy the $D >24$\,mJy cut gives an $S_{250} < 32.6$\,mJy upper limit, and since the typical $S_{500}/S_{250}$ ratio of our catalogue is 1.55, using this limit will give a better overlap between our sample and the \citet{dowell2014} catalogue. Integrating our power law for $S_{500} > 40$\,mJy results in 2.8 objects per deg$^2$, which is closer to the \citet{dowell2014} result.

\section{Colours and SED fits}
\label{sec:tobs}

\begin{figure*}
\centering
\includegraphics[width=0.8\textwidth]{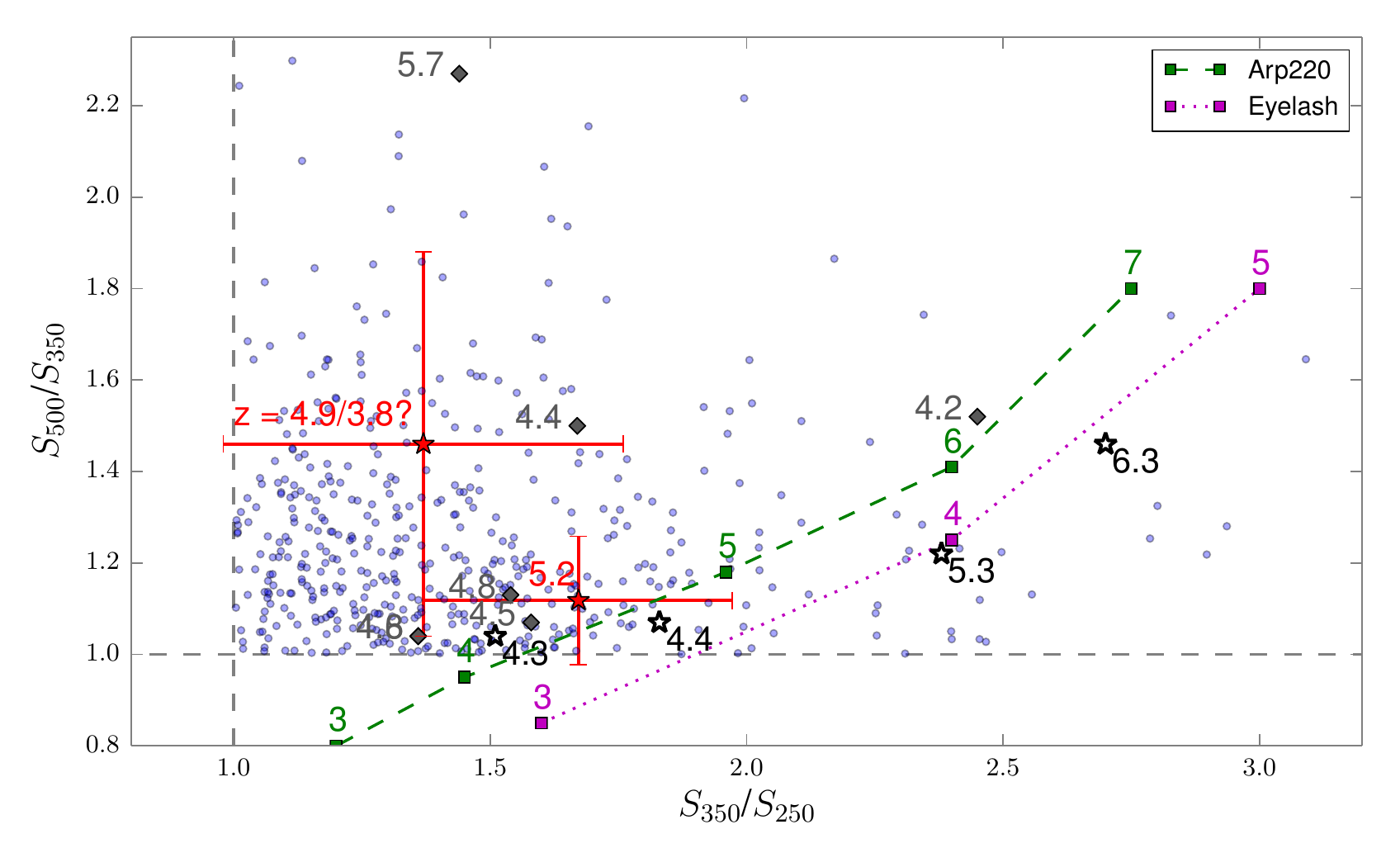} 
\caption{SPIRE colour-colour plot of red sources. Blue filled circles represent the objects in the current catalogue. The green dashed line represents the redshift-track for the starburst galaxy Arp220 \protect\citep{rangwala2011} and the dotted purple line for SMMJ21352--102 (Cosmic Eyelash, \protect\citealt{swinbank2010}), with the labels on these curves showing the redshifts. Red filled stars show two sources from the catalogue discussed in our paper that have ALMA redshift measurements (see Section~\ref{sec:alma}). These are, from left to right: HELMS{\_}RED{\_}31 with $z = 3.798$ ; and HELMS{\_}RED{\_}4 with $z = 5.162$. For these two sources we plot the errors in the colour measurements. These uncertainties are representative of the typical errors of the colours of sources in our catalogue. The black open star symbols represent four sources from the \protect\citet{dowell2014} red source sample that have spectroscopic redshift measurements. From left to right: FLS1 ($z = 4.3$); FLS5 $(z = 4.4$); LSW102 ($z = 5.3$); and FLS3 ($z = 6.3$). Grey diamonds represent spectroscopically confirmed $z>4$ sources from the SPT lensed galaxy sample \protect\citep{weiss2013} that also have red SPIRE colours. \label{fig:colourcolour}}
\end{figure*}

The interpretation of our red colour-selection as leading to a catalogue rich in high-\textit{z} galaxies relies on the assumption that galaxies with rising flux densities towards longer wavelengths are at high redshifts. Although the temperature of starburst galaxies could be rising slightly towards high $z$, it does not rise as fast as $(1+z)$, and the observed-frame temperature $T_{\text{obs}} = T_\text{d}/(1+z)$ drops with redshift. This behaviour is evident in the catalogue of 25 strongly-lensed galaxies selected at 1.4\,mm with the South Pole Telescope \citep{weiss2013,vieira2013}, where the apparent temperature is fit by $T_{\text{obs}}= [11.1 - 0.8 (1+z)]$\,K. This pattern is consistent with examining the spectral energy distributions (SEDs) of several lower redshift dusty starburst galaxies, and measuring their expected flux-ratios at the SPIRE wavelengths for different redshifts. As we noted before, not all high-$z$ galaxies are 500-$\mu$m-risers \citep[e.g.][]{Rowan-Robinson2014, riechers2014,smolcic2015}, but we assume that the galaxies we select this way are predominantly at $z >4$ without significant contamination from lower redshift objects.  

In Fig.~\ref{fig:colourcolour} we present the colour-colour plot for our objects. As a comparison we also show the redshift tracks for two starburst galaxies, as well as the colours of our two red sources that have ALMA redshift measurements and four  $z >4$ galaxies from the \citet{dowell2014} sample. Most of the galaxies from the SPT lensed galaxy sample \citep{weiss2013} that have spectroscopically confirmed $z>4$ redshifts also have red SPIRE colours. We also show these sources in our colour-colour plot. The error bars on the two sources with known ALMA redshift are representative of the typical errors for the sources in our catalogue. Due to the large photometric uncertainties the measured colours have a very large scatter. Blending will also change the observed colours. As discussed before, we included the blends in our catalogue, assuming that they contain at least one red source, and we addressed the flux boosting effect caused by blending in our number counts simulation. However, when investigating individual sources in the catalogue we have to be careful how we interpret our flux density values. The correlation values included in the catalogue and also visual inspection of the selected sources in the higher resolution SPIRE bands can help identifying sources that have less reliable photometry due to blending effects.

\begin{figure}
\centering 
\includegraphics{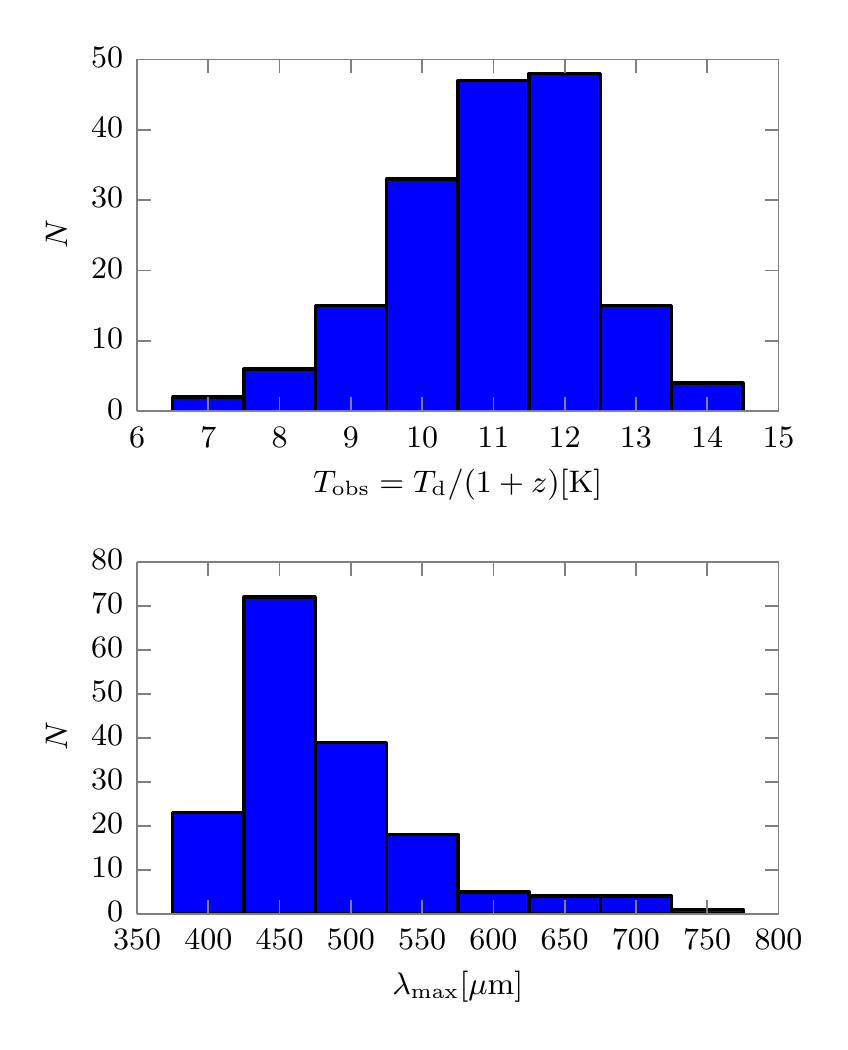} 
\caption{Distribution of the observed temperature $T_{\text{d}}/(1+z)$ and the observed SED peak wavelength $\lambda_{\text{max}}$ of our red sources, measured by fitting an optically thick modified blackbody spectrum to our SPIRE flux densities, as described in Section~\ref{sec:tobs}.\label{fig:zfit}}
\end{figure}

Similarly to \citet{dowell2014} we  fit our observed SPIRE flux densities with an optically thick modified blackbody spectrum:
\begin{equation}
S_{\nu} = \Omega \times [1-\text{exp}(-(\nu/\nu_0)^\beta)] \times B_{\nu}(T).
\end{equation}
Here $B_{\nu}(T)$ is the Planck function, $\Omega$ is the solid angle of the source, and $\nu_0$ is the frequency where the optical depth is unity. Redshifting a thermal SED has a similar effect on the observed submillimetre colours as changing the dust temperature, and using the SPIRE fluxes alone we can only measure the combination $T_{\text{obs}} = T_{\text{dust}}/(1+z)$. We use the affine invariant MCMC code described in \citet{dowell2014}, while marginalizing over the Gaussian priors $\beta = 1.8 \pm 0.3$ and $\lambda_0 (1+z) = (1100 \pm 400)$\,$\mu$m. The resulting  observed-frame temperature distribution for our isolated sources is plotted in Fig.~\ref{fig:zfit}. We also show the distribution of $\lambda_{\text{max}}$, the observed wavelength where the SED peaks. 

The mean observed temperature is $(11.03 \pm 1.91)$\,K. As expected from high-redshift galaxies, this is cooler than $T_{\text{obs}}$ for SPIRE-selected galaxies in general \citep[e.g.][]{amblard2010,casey2012}. However, we have to be very careful when comparing temperature values quoted in the literature. If we apply an optically thin SED model, where $\lambda_0 \rightarrow 0$, the fitted $T_{\text{obs}}$ values can decrease by about $15 \%$, so the observed temperatures can have a different meaning depending on the specific SED model applied. Using the same SED model as \citet{dowell2014}, our observed temperature distribution is similar to their measurements, showing that we select a similar population of sources in our maps. Red sources with known redshifts have a very warm inferred intrinsic dust temperature. 

The three sources with spectroscopic redshift estimations listed in \citet{dowell2014}, FLS1 ($z=4.29$), FLS5 ($z=4.44$) and LSW20 ($z=3.36$), have dust temperatures of 63\,K, 59\,K and 48\,K, respectively, and \citet{riechers2013} quote $T_\text{d} = 56$\,K for the $z = 6.3$ source HFLS3. The observed temperatures of these sources are 11\,K, 10\,K, 9\,K and 8\,K, respectively. The observed SED peak wavelength $\lambda_{\text{max}}$ is more directly constrained by the data than $T_{\text{obs}}$.  For a sample with a similar $\lambda_{\text{max}}$ distribution as the one we measure (bottom panel in Fig.~\ref{fig:zfit}), \citet{dowell2014} estimated a mean photometric redshift of $z = 4.7$ by determining priors of the rest-frame peak wavelength, based on different comparison samples. While we did not carry out a similar analysis, based on the similar selection function, and the similar measured $\lambda_{\text{max}}$ and $T_{\text{obs}}$ distribution of our sample, we can assume that our catalogue also consists of mostly high-redshift objects.

\section{Follow-up results}
\label{sec:followup}

Previous follow-up observations of red sources have already shown that we can successfully select high-redshift dusty galaxies based on their red SPIRE colours, but to confirm that this is true for our whole sample more observations will be needed. Here we summarize spectroscopic redshift measurements with ALMA and millimetre-wave photometric follow-up measurements with CSO/MUSIC for a sub-sample of the red sources in the HeLMS catalogue. 

\subsection{ALMA spectroscopy}
\label{sec:alma}

\begin{figure*}
\centering
\includegraphics{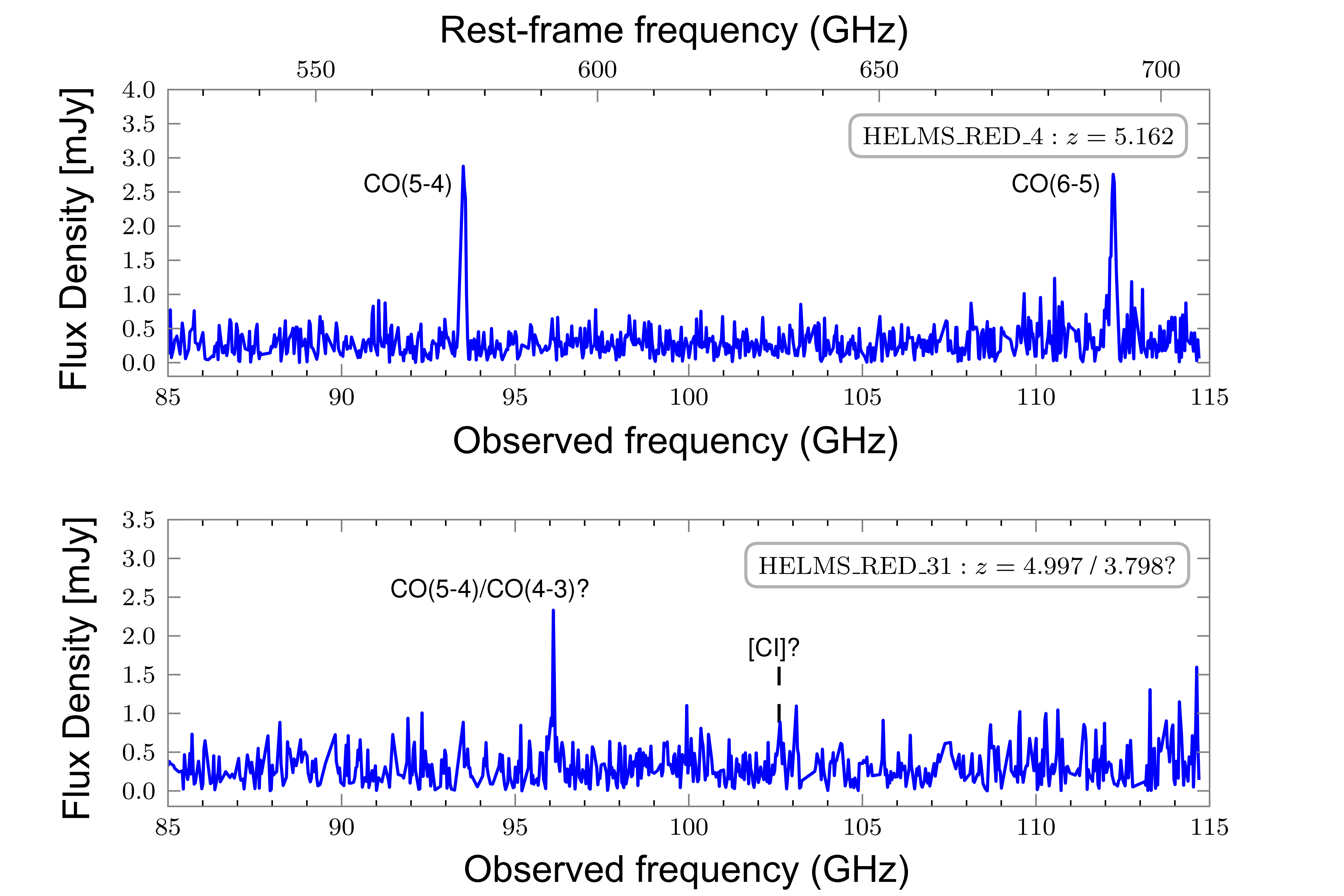} 
\caption{ALMA spectra for two red sources in our catalogue. In the spectrum of HELMS{\_}RED{\_}4 we detect two CO lines and their observed frequencies correspond to redshift $z=5.162$. HELMS{\_}RED{\_}31 has only one high signal-to-noise line in its spectrum and the redshift can be either $z=4.997$ or $z=3.798$, depending on whether the low signal-to-noise ratio spectral feature at 102.6\,GHz is only a noise fluctuation or a faint [CI] spectral line. See the discussion in Section~\ref{sec:alma} for details.\label{fig:almaspec}. We note that recent VLA observations confirm the lower $z=3.798$ value (Riechers et al. in prep.).}
\end{figure*}

We carried out spectroscopic observations for two of our red sources, HELMS{\_}RED{\_}4 and HELMS{\_}RED{\_}31, using the Atacama Large Millimeter/submillimeter Array (ALMA) during the Cycle 2 operational phase, to determine their redshifts. These two objects have photometric redshift estimates of $z_{\text{phot}}=5.27 \pm 0.12$ and $z_{\text{phot}}=5.28 \pm 0.27$, respectively, based on the SPIRE flux densities and additional millimetre-wave interferometric follow-up observations (the photo-$z$ estimation method will be presented in Clements et al. in prep.). The observations were carried out in Band 3, covering frequencies between 84 and 116\,GHz, which contains the redshifted CO rotational lines typically up to the $J=6-5$ transition at the expected redshifts. 

The observed spectra are shown in Fig. ~\ref{fig:almaspec}. In the spectrum of HELMS{\_}RED{\_}4 we detect two lines, unambiguously identified as the CO($5-4$) and CO($6-5$) transitions. These correspond to a redshift of $z=5.162$, which is in very good agreement with the photo-\textit{z} estimate of $5.27$.  Only a single strong line is seen in the spectrum of HELMS{\_}RED{\_}31, and there are several spectral lines that could fall into this region. We can discard the possibility that the observed line is CO($6-5$) or a higher transition, since then we would always detect other CO rotational lines in the observed spectrum. If we identify the detected line as the CO($5-4$) transition then the redshift of this source is $z=4.997$  and we would not see any other lines in the observed spectral range; this is consistent with the  photo-\textit{z}  estimate. If, on the other hand, the detected line corresponds to the CO($4-3$) transition then the redshift is $z=3.798$ and we should be able to detect a [CI] line at 102.6\,GHz. There is a possible low signal-to-noise peak near to this frequency, but the spectral feature in question is not larger than a dozen or so other spikes in the observed spectrum, and hence the identification is not definite from this measurement alone. We note that recent observations with the Karl G. Jansky Very Large Array (VLA) confirm the $z=3.798$ redshift value (Riechers et al. in prep.).

Based on the redshift measurement of $z = 5.162$ and the observed temperature $T_{\text{obs}} = 10.9 \pm 0.9$\,K, we infer that HELMS{\_}RED{\_}4 has a dust temperature of $T_\text{d} = 67 \pm 6$\,K, which is similar to the higher than average dust temperatures of other red sources mentioned in Section~\ref{sec:tobs}. These follow-up results all strengthen the hypothesis that red SPIRE colours select mainly high-redshift galaxies that contain warmer dust, instead of very low redshift cold objects.  

\subsection{CSO/MUSIC millimeter-wave observations}
\label{sec:music}

We observed four of our 500-micron-bright sources (HELMS{\_}RED{\_}3, HELMS{\_}RED{\_}4, HELMS{\_}RED{\_}6 and HELMS{\_}RED{\_}7) at the Caltech Submillimeter Observatory using the Multiwavelength Submillimeter Inductance Camera \citep[MUSIC,][]{sayers2014}. MUSIC observes simultaneously in four bands centred on frequencies of 143, 213, 272, and 326\,GHz (2.09, 1.4, 1.1 and 0.92\,mm in wavelength), with PSFs that have FWHMs of $48''$, $ 36''$, $32''$, and $29''$, respectively. We note that these PSFs differ from the ones described in \citet{sayers2014} due to a change in the optical configuration of the instrument. The 326\,GHz observing band overlaps with an optically thick absorption feature in the atmosphere, and as a result its sensitivity is significantly degraded relative to the other three observing bands. The results of the measurements are listed in Table ~\ref{tbl:music}. 

\begin{table}
\caption{MUSIC flux density measurements at 143, 213, 272, and 326\,GHz.\label{tbl:music}}
\begin{tabular}{lcccc}
\hline
Source & $S_{\text{326}}$ & $S_{\text{272}}$ &$ S_{\text{213}}$ &$ S_{\text{143}}$ \\
 & [mJy] &  [mJy] & [mJy] & [mJy]\\
\hline
\scriptsize HELMS{\_}RED{\_}3 & $100.3 \pm 52.8$ & $24.9 \pm 12.3$ & $16.0 \pm 6.9$  & $-2.5 \pm 7.7$\\ 
\scriptsize HELMS{\_}RED{\_}4 & $\phantom{1}65.2 \pm 57.3$ & $32.8 \pm 12.4$ & $19.4 \pm 6.7$  & $\phantom{1} \,\, 8.4 \pm 7.4$\\ 
\scriptsize HELMS{\_}RED{\_}6 & $-5.6 \,\, \pm 37.6$ & $32.3 \pm \phantom{1}9.8$ & $\phantom{1}8.8 \pm 4.3$  & $ \,\, 13.1 \pm 5.7$\\
\scriptsize HELMS{\_}RED{\_}7 & $108.7 \pm 35.1$ & $62.4 \pm 13.5$ & $18.1 \pm 5.1$ & $\phantom{1} \,\, 1.8 \pm 7.1$\\
\hline
\end{tabular} 
\end{table}

\begin{figure*}
\centering
\includegraphics{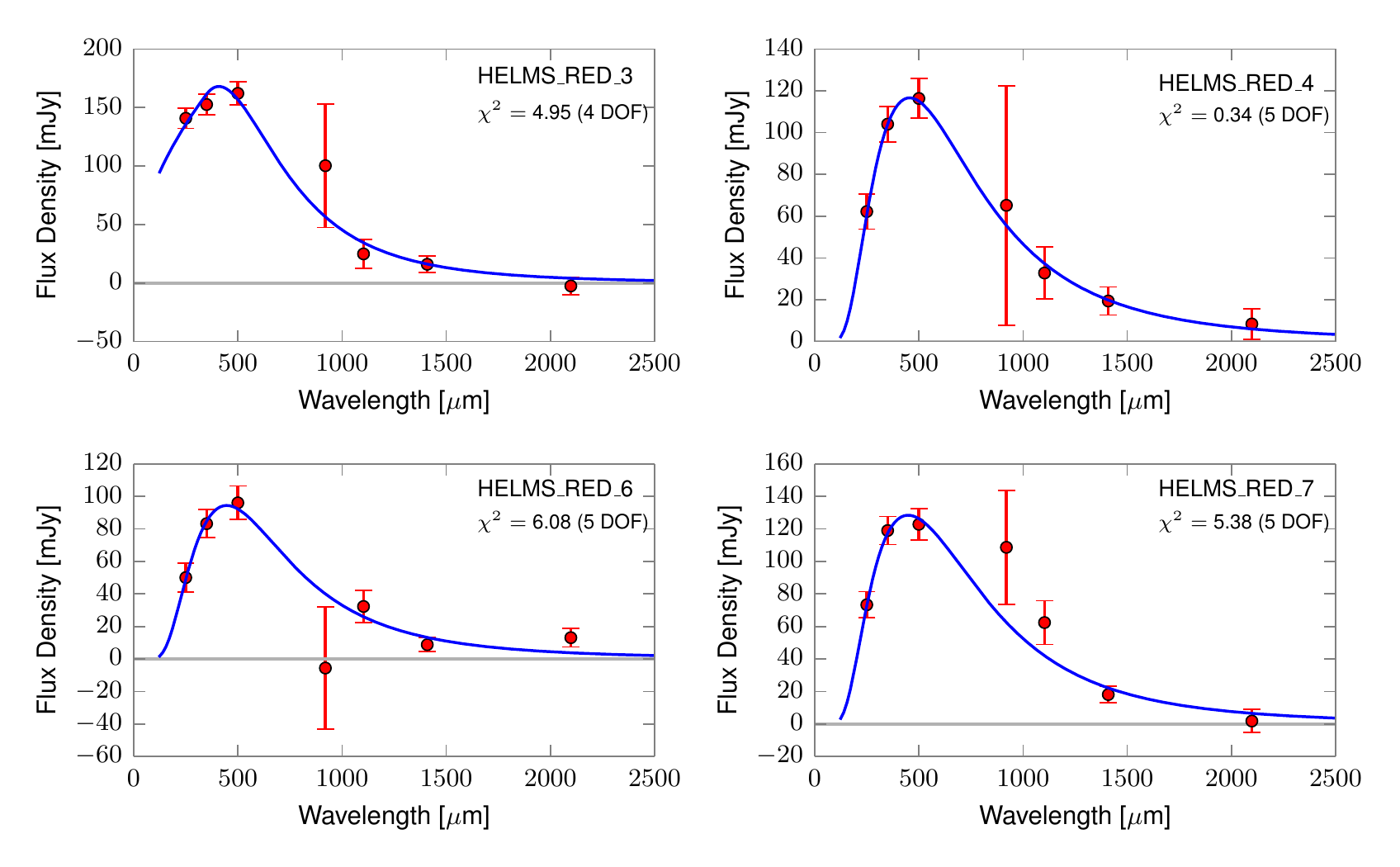} 
\caption{Modified blackbody SED fits to the measured SPIRE and MUSIC flux densities as discussed in Section~\ref{sec:music}. For each object we give the $\chi^2$ for the best fit, as well as the number of degrees of freedom. \label{fig:musicfits}}
\end{figure*}

In Fig.~\ref{fig:musicfits} we show modified blackbody fits to the SPIRE and MUSIC photometry measurements for these four sources. We apply Gaussian priors on $\beta$ and $\lambda_0(1+z)$, as discussed in Section~\ref{sec:tobs}, thus we effectively fit for two parameters, $T_\text{obs}$ and $S_{500}$ (except in the case of HELMS{\_}RED{\_}3, where attaching an $S_\nu \propto \nu^{-\alpha}$ power law onto the short wavelength side of the peak results in a better fit to our flux densities). The $\chi^2 $ values are acceptable for all fits, and the observed temperatures are $11.10 \pm 1.18 \text{\,K}, 10.85 \pm 0.95 \text{\,K}, 10.79 \pm 1.32 \text{\,K} \text{ and } 11.28 \pm 0.80 \text{\,K}$ for HELMS{\_}RED{\_}3, HELMS{\_}RED{\_}4, HELMS{\_}RED{\_}6, and HELMS{\_}RED{\_}7, respectively. Using the SPIRE flux densities only we obtain $11.89 \pm 1.16 \text{\,K}, 10.85 \pm 0.96 \text{\,K}, 10.70 \pm 1.16 \text{\,K} \text{ and } 11.33 \pm 0.90 \text{\,K}$ observed temperatures from the fits, confirming that fitting to the SPIRE flux densities alone and using the above described Gaussian priors on $\beta$ and  $\lambda_{0,\text{obs}}$ works well to describe the thermal SED.

\section{Conclusions}

Using a map-based search technique we have created a catalogue of 477 sources selected in the HeLMS field with flux densities $S_{500} > S_{350} > S_{250}$ and a $5\sigma$ cut-off $S_{500} > 52$\,mJy at 500\,$\mu$m. We discarded cosmic rays from the catalogue and flagged radio sources that can have similar SPIRE flux ratios to dusty galaxies at high redshift. 

We measured the raw number counts of our sample and used Monte Carlo simulations to infer the possible intrinsic number counts, taking into account corrections for completeness, Eddington bias, false detections and blending effects. Similarly to \citet{dowell2014} we have found an excess of red sources above the numbers predicted by galaxy evolution models that best describe the total $Herschel$ number counts. Our sample is, however, much larger than the \citet{dowell2014} sample and we are now able to measure the slope of the differential number counts: the 500\,$\mu$m counts decrease steeply towards higher flux densities. Since we select sources at the 500\,$\mu$m resolution, blending is a significant contaminating effect. Our data-set and the available ancillary data is not optimal for finding higher resolution counterparts for deblending, so we did not remove the blends from our catalogue, but instead we took their effect into account in the simulations. 

Our SED fit results show a similar observed temperature and peak wavelength distribution as in the \citet{dowell2014} sample, suggesting that we are detecting the same population of sources. We also presented ALMA follow-up observations of two sources, which further proved the efficiency of selecting high-redshift starbursts using this approach. One of our sources is at redshift $z = 5.162$ and while the redshift determination for the other source is not unambiguous from the ALMA spectrum alone, recent VLA observations confirmed a redshift of $z = 3.798$. We additionally presented CSO/MUSIC photometric measurements that can constrain the long wavelength side of the SED and have found consistent results between the SPIRE flux densities and the MUSIC photometry points.

\section*{Acknowledgements}

AC acknowledges support from the National Aeronautics and Space Administration under Grant No. 12-ADAP12-0139 issued through the ADAP programme.EI acknowledges funding from CONICYT/FONDECYT postdoctoral project N$^\circ$:3130504. JLW is supported by a European Union COFUND/Durham Junior Research Fellowship under EU grant agreement number 267209, and acknowledges additional support from STFC (ST/L00075X/1). RJI acknowledges support from the European Research Council in the form of the Advanced Investigator Program, 321302, COSMICISM. SO acknowledges support from the
Science  and  Technology  Facilities  Council  (grant  number ST/L000652/1).SRS acknowledges support from NASA NESSF. The construction of MUSIC was supported by JPL RTD, NASA APRA, NSF ATI and NSF AAG. The Dark Cosmology Centre is funded by the Danish National Research Foundation. This paper makes use of the following ALMA data: ADS/JAO.ALMA\#2013.1.00449.S. ALMA is a partnership of ESO (representing its member states), NSF (USA) and NINS (Japan), together with NRC (Canada), NSC and ASIAA (Taiwan), and KASI (Republic of Korea), in cooperation with the Republic of Chile. The Joint ALMA Observatory is operated by ESO, AUI/NRAO and NAOJ. SPIRE  has  been developed by a consortium of institutes led by Cardiff Univ. (UK) and including Univ. Lethbridge (Canada); NAOC (China); CEA, LAM (France); IFSI, Univ. Padua (Italy); IAC (Spain); Stockholm Observatory (Sweden);Imperial College London,  RAL,  UCL-MSSL, UKATC, Univ.  Sussex (UK); Caltech, JPL, NHSC, Univ.  Colorado  (USA).  This  development  has  been  supported by  national  funding  agencies:   CSA  (Canada);  NAOC (China);  CEA,  CNES,  CNRS  (France);  ASI  (Italy); MCINN  (Spain);  SNSB  (Sweden);  STFC  (UK);  and NASA  (USA). HCSS / HSpot / HIPE are joint developments by the Herschel  Science  Ground  Segment  Consortium, consisting of ESA, the NASA Herschel Science Center, and the HIFI, PACS and SPIRE consortia. This  research  has  made  use  of  data  from  the  HerMES project (http://hermes.sussex.ac.uk/).  HerMES is a Herschel Key Programme utilizing Guaranteed Time from the SPIRE instrument team, ESAC scientists and a mission scientist. HerMES is described in Oliver et al. (2012). The data presented in this paper will be released through  the  HerMES  Database  in  Marseille,  HeDaM (http://hedam.lam.fr/HerMES/).


\bibliographystyle{mnras}
\bibliography{v_asboth_manuscript}

\bsp	
\label{lastpage}
\end{document}